\documentclass[prl,letterpaper,twocolumn,nofootinbib,preprintnumbers,superscriptaddress]{revtex4-1} 

\pdfpageattr {/Group << /S /Transparency /I true /CS /DeviceRGB>>}

\usepackage[utf8]{inputenc}
\usepackage{newtxtext,newtxmath}

\usepackage{titlesec}
\usepackage{dcolumn}
\usepackage{bm}
\usepackage{amsmath,amssymb,amsfonts}

\usepackage{graphicx}

\usepackage{color}
\usepackage{bbold}

\usepackage{slashed}
\usepackage[svgnames]{xcolor}
\usepackage[english]{babel}
\usepackage{blindtext}
\usepackage{microtype}
\usepackage{tikz}

\usepackage[colorlinks=true,linkcolor=blue,urlcolor=blue,citecolor=blue]{hyperref}
\usepackage{ifpdf}

\newcommand{\Eq}[1]{Eq.~\eqref{#1}}
\newcommand{\Eqs}[1]{Eqs.~(\ref{#1})}

\newcommand{\Fig}[1]{Fig.~{\ref{#1}}}

\newcommand{\vect}[1]{\boldsymbol{#1}}

\newcommand{\sh}[1]{\slashed{#1}}

\graphicspath{{.}{figures/}} 

\def\hs{\hspace}

\def\no{\nonumber}

\def\lf{\left}
\def\rg{\right}

\makeatletter
\newcommand*\if@single[3]{%
  \setbox0\hbox{${\mathaccent"0362{#1}}^H$}%
  \setbox2\hbox{${\mathaccent"0362{\kern0pt#1}}^H$}%
  \ifdim\ht0=\ht2 #3\else #2\fi
  }
\newcommand*\rel@kern[1]{\kern#1\dimexpr\macc@kerna}
\newcommand*\widebar[1]{\@ifnextchar^{{\wide@bar{#1}{0}}}{\wide@bar{#1}{1}}}
\newcommand*\wide@bar[2]{\if@single{#1}{\wide@bar@{#1}{#2}{1}}{\wide@bar@{#1}{#2}{2}}}
\newcommand*\wide@bar@[3]{%
  \begingroup
  \def\mathaccent##1##2{%
    \if#32 \let\macc@nucleus\first@char \fi
    \setbox\z@\hbox{$\macc@style{\macc@nucleus}_{}$}%
    \setbox\tw@\hbox{$\macc@style{\macc@nucleus}{}_{}$}%
    \dimen@\wd\tw@
    \advance\dimen@-\wd\z@
    \divide\dimen@ 3
    \@tempdima\wd\tw@
    \advance\@tempdima-\scriptspace
    \divide\@tempdima 10
    \advance\dimen@-\@tempdima
    \ifdim\dimen@>\z@ \dimen@0pt\fi
    \rel@kern{0.6}\kern-\dimen@
    \if#31
      \overline{\rel@kern{-0.6}\kern\dimen@\macc@nucleus\rel@kern{0.4}\kern\dimen@}%
      \advance\dimen@0.4\dimexpr\macc@kerna
      \let\final@kern#2%
      \ifdim\dimen@<\z@ \let\final@kern1\fi
      \if\final@kern1 \kern-\dimen@\fi
    \else
      \overline{\rel@kern{-0.6}\kern\dimen@#1}%
    \fi
  }%
  \macc@depth\@ne
  \let\math@bgroup\@empty \let\math@egroup\macc@set@skewchar
  \mathsurround\z@ \frozen@everymath{\mathgroup\macc@group\relax}%
  \macc@set@skewchar\relax
  \let\mathaccentV\macc@nested@a
  \if#31
    \macc@nested@a\relax111{#1}%
  \else
    \def\gobble@till@marker##1\endmarker{}%
    \futurelet\first@char\gobble@till@marker#1\endmarker
    \ifcat\noexpand\first@char A\else
      \def\first@char{}%
    \fi
    \macc@nested@a\relax111{\first@char}%
  \fi
  \endgroup
}
\makeatother

\begin{document}

\title{Nucleon Quark Distribution Functions from the Dyson--Schwinger Equations}

\author{Kyle D. Bednar}
\affiliation{Center for Nuclear Research, Department of Physics, Kent State University, Kent OH 44242 USA}

\author{Ian C. Clo\"et}
\affiliation{Physics Division, Argonne National Laboratory, Argonne, IL 60439 USA}

\author{Peter~C.~Tandy}
\affiliation{Center for Nuclear Research, Department of Physics, Kent State University, Kent OH 44242 USA}

\affiliation{CSSM, Department of Physics, University of Adelaide, Adelaide SA 5005, Australia}

\preprint{ADP-18-7/T1055}

\begin{abstract}
We present results for the nucleon's leading-twist spin-independent valence parton distribution functions obtained from a theoretical framework based on the Dyson-Schwinger equations (DSEs) of QCD that previously gave an excellent description of nucleon electromagnetic form factors. We employ the rainbow-ladder truncation of the DSEs and utilize nucleon bound state amplitudes from  the Poincar\'e-covariant Faddeev equation, where the dominant scalar and axial-vector quark-quark correlations are included. This DSE framework is used to numerically evaluate the first 20 moments of the valence $u$ and $d$ quark distribution functions, from which the $x$-dependence of the distributions is found to be well constrained. We find good agreement with empirical parameterizations of experimental data and make the prediction that the $d/u$ ratio in the $x\to 1$ limit, invariant under scale evolution, takes the value $d/u \to 0.087 \pm 0.010$.  We find that this ratio is rather sensitive to the strength of axial-vector diquark correlations. However, contrary to a naive expectation, our result for the $d/u$ ratio in the $x\to 1$ limit does not vanish when only scalar diquark correlations are present, although it is an order of magnitude smaller than our $d/u$ result that also includes axial-vector diquarks.  The valence quark distribution results are set in a broader context via a simple pion cloud model estimate of sea-quark light-cone momenta and gluon light-cone momentum.  
\end{abstract}

\maketitle
\noindent\textbf{Introduction:} 
The ongoing quest to measure and understand the partonic structure of hadrons, {\it e.g.}, via their electromagnetic form factors and parton distribution functions (PDFs), will deepen our understanding of nonperturbative quantum chromodynamics (QCD) and shed light on key emergent phenomena such as dynamical chiral symmetry breaking (DCSB) and color confinement. Form factors and PDFs are amalgams of short-distance perturbative processes and nonperturbative dynamics, and they challenge our understanding of both aspects and their interface. The calculation of these observables in a framework with a well-defined connection to QCD -- over all kinematic domains -- has been a long-standing challenge. For example, in the calculation of spacelike electromagnetic form factors, lattice implementations of QCD have until recently been limited to momentum transfers in the range $0 \leqslant Q^2 \lesssim 1\,$GeV$^2$~\cite{Green:2014xba}, and  despite new techniques that make calculations up to $Q^2 \sim 6\,$GeV$^2$ possible~\cite{Chambers:2017tuf,Koponen:2017fvm}, form factors at the significantly higher $Q^2$ values needed to explore the transition to perturbative QCD predictions~\cite{Farrar:1979aw,Lepage:1979zb} remain out of reach.   With regard to PDFs, standard lattice QCD techniques are presently limited to the first few moments~\cite{Lin:2017snn}, although significant progress is being made through recently proposed spacelike correlator approaches~\cite{Ji:2013dva,Xiong:2013bka,Ji:2014gla,Lin:2014zya,Alexandrou:2015rja,Radyushkin:2017cyf,Ma:2017pxb} to directly infer PDF behavior as a function of the intrinsic light-cone momentum fraction variable $x$. Nonetheless, the domains $x \lesssim 0.2$ and $x \gtrsim 0.8$  will likely remain a practical challenge for some time due to difficulties in achieving large volumes and large hadron momenta  on the lattice.

In this work we utilize the Dyson--Schwinger equations (DSEs) of QCD to calculate the leading-twist spin-independent quark distribution functions of the nucleon. The DSE framework incorporates numerous key features of QCD, {\it e.g.}, DCSB, quark confinement, a running quark mass, and it is applicable over all kinematic domains and momentum scales since it is Poincar\'e-covariant and displays asymptotic freedom. The DSE approach used here is the same as that employed in Ref.~\cite{Cloet:2008re} to calculate and predict the nucleon's electromagnetic form factors over the range $0 \leqslant Q^2 \lesssim 20\,$GeV$^2$, namely, the rainbow-ladder truncation of QCD's DSEs, which accounts for the momentum dependence of propagators generated by gluon exchange. In other works the DSEs have been used with success to determine, {\it e.g.}, the pion's elastic and transition form factors over all spacelike momenta~\cite{Chang:2013nia,Raya:2015gva}, pion and kaon distribution amplitudes~\cite{Chang:2013pq,Shi:2014uwa} and PDFs~\cite{Nguyen:2011jy,Chang:2014lva}, and a pion generalized parton distribution~\cite{Mezrag:2014jka}. 

Use of the DSE description of nucleon structure obtained in Ref.~\cite{Cloet:2008re} to also determine the nucleon's valence parton distribution functions establishes observables from a single framework for both electromagnetic form factors and PDFs. This enables inferences to be made with regard to the core features of the DSE approach to the baryon sector~\cite{Eichmann:2016yit} and  the robustness of associated approximations. This is essential in the baryon sector because some of the simplicities afforded by meson calculations, as bound states of two dressed-quarks, are not available for the three-quark sector.

\medskip
\noindent\textbf{DSE Framework for Nucleon PDFs:} 
In the Bjorken kinematical limit, the leading-twist spin-independent quark distribution functions for a quark of flavor $q$ are defined by the explicitly Poincar\'e-invariant matrix element~\cite{Jaffe:1996zw,Diehl:2003ny}
\begin{align}
q(x) = \int \frac{d\lambda}{4\,\pi}\  e^{-i x\, P\cdot n\, \lambda} \,
\left< P\left|\bar{\psi}_q(\lambda n)\,\sh{n}\,\psi_q(0)\right|P\right>_c
\label{eq:PDF_defn}
\end{align}
in Minkowski metric, with the  light-like longitudinal vector given by $n^\mu=(1,\,\vect{0}_T,\,-1)$ in the target rest frame, and with the nucleon Dirac spinor having normalization $\bar{u}(P)\,u(P) = 2\,m_N$. The DSE approach is based on equations that couple $n$-point Green functions and does not easily accommodate the Wilson line that needs to be included in Eq.~\eqref{eq:PDF_defn} to enforce strict color gauge invariance. In  light-cone gauge ($n\cdot A=0$) the Wilson line is formally unity~\cite{Jaffe:1996zw}; however, a practical implementation of the DSE approach has historically implied the choice of Landau gauge.  As this work requires a series of results for nonperturbative elements such as quark and gluon propagators, quark-quark correlation amplitudes, and nucleon bound state amplitudes, we adopt these elements from previous studies which have provided good descriptions of many meson and nucleon properties~\cite{Cloet:2013jya}, including confirmed predictions for nucleon electromagnetic form factors~\cite{Cloet:2008re,Riordan:2010id}. We therefore postpone the challenge of a rigorous treatment of the Wilson line, noting that in the DSE framework a Wilson line can be formulated as a $t$-matrix equation and that in at least one estimate~\cite{Polyakov:1997ea} the numerical impact on leading-twist PDFs was found to be small.

\begin{figure}[tbp]
\centering\includegraphics[width=\columnwidth]{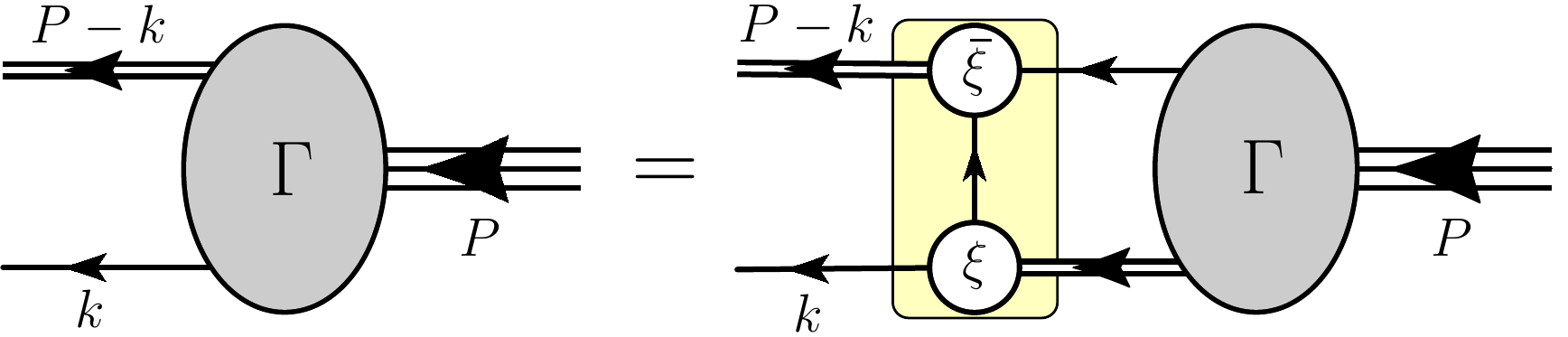}
\caption{The Poincar\'e-covariant Faddeev equation which describes the nucleon as a bound state of three dressed quarks. Here, the quark-diquark approximation is made explicit, where the single line represents a quark propagator and the double-line a diquark propagator. The object $\Gamma$ is the homogeneous Faddeev vertex, $\xi$ is the diquark Bethe-Salpeter vertex, and $\protect\widebar{\xi}$ is its conjugate. The shaded rectangle is the Faddeev kernel.}
\label{fig:faddeev}  
\end{figure}

We use the solution of the Poincar\'e-covariant Faddeev equation~\cite{Cahill:1988dx,Eichmann:2009qa} 
in the form obtained in Ref.~\cite{Cloet:2008re} to describe nucleon structure.   In principle this integral equation sums all possible interactions among the three dressed quarks, including irreducible 3-quark interactions. The inherent numerical problems are rendered tractable through limitation of the Faddeev kernel to two-body interactions, use of the rainbow-ladder truncation of QCD dynamics, and then approximating the quark-quark scattering $t$-matrix in the separable form~\cite{Oettel:2001kd,Cloet:2008re}
\begin{align}
t_D(\ell,\ell',K) = \xi_D(\ell)\, \tau_D(K) \, \bar{\xi}_D(\ell'),
\label{eq:sep_t}
\end{align}   
where $\xi_D(\ell)$ is the quark-quark (diquark) homogeneous Bethe-Salpeter amplitude for a diquark of type $D$ with relative momentum $\ell$ between the quarks, $\bar{\xi}_D(\ell')$  is the conjugate amplitude, and $\tau_D(K)$ is the effective diquark propagator with momentum $K$.  Such a separable approximation is of long-standing utility~\cite{Cahill:1988dx,Oettel:2001kd} and has been shown to capture the essential physics of  a number of baryon observables~\cite{Cloet:2014rja,Segovia:2015hra,Eichmann:2016yit}.  The Faddeev equation, with the quark-diquark approximation made explicit, is illustrated in Fig.~\ref{fig:faddeev}. In this representation the  momentum-dependent Bethe-Salpeter amplitudes $\xi_D(\ell)$ and $\widebar{\xi}_D(\ell)$ influence the dissociation and rearrangement of correlated quark pairs in the various diquark channels $D$.  More detailed forms of these quantities and of the Faddeev amplitudes are discussed later and given in Ref.~\cite{Cloet:2008re}.

\begin{figure}[tbp]
\centering\includegraphics[width=0.68\columnwidth]{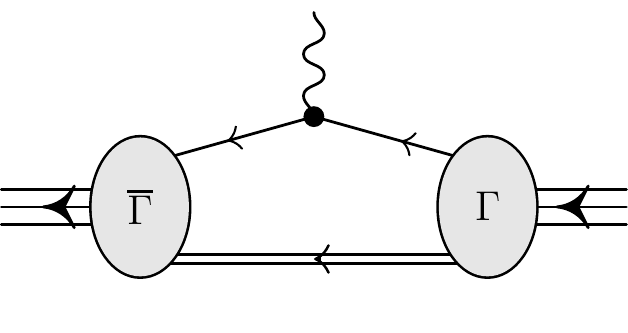}
\centering\includegraphics[width=0.68\columnwidth]{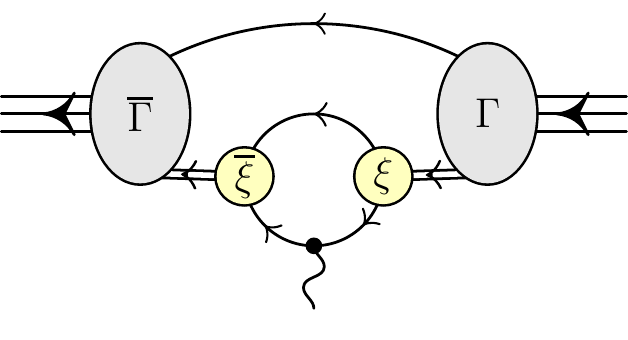}
\centering\includegraphics[width=0.68\columnwidth]{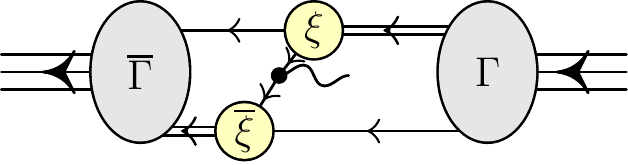}
\caption{The three dominant diagrammatic mechanisms for the nucleon's quark distribution functions when the bound state dynamics is described by a quark-diquark approximation. The elementary operator insertion on a quark line, as described in the text, is represented as a wavy line.  The double line is the diquark propagator, the small shaded circles are the diquark  Bethe-Salpeter amplitudes, and the larger shaded ovals are the nucleon's Faddeev amplitudes. {\it Top panel}:  illustrates diagram 1 which is described by \protect\Eq{eq:n_moms_diag1}.   {\it Center panel}:  illustrates diagram 2 which is described by \protect\Eqs{eq:moms_diquarks}, \protect\eqref{N_moms_Diag2} and \protect\eqref{Diag2_prod}.  {\it Bottom panel}: illustrates diagram 3 which is described by \protect\Eq{eq:n_moms_diag3}.  }
\label{fig:Diag123}  
\end{figure}

Through use of the rainbow-ladder DSE truncation and the consequent nucleon Faddeev amplitudes of the quark-diquark type, the leading-twist valence quark distributions defined by Eq.~\eqref{eq:PDF_defn} can be represented by the three diagrams illustrated in Fig.~\ref{fig:Diag123}.    Here the operator insertion (wavy line) on each quark line having momentum $k$ has the form $\delta\lf(k\cdot n - x\,P\cdot n\rg)\,\sh{n}$, with $P$ the 4-momentum of the nucleon.\footnote{In principle two other diagrams appear, the so-called seagull terms~\cite{Cloet:2008re}, in which the operator insertion is on a quark interior to a diquark correlation amplitude just before or after a quark is exchanged with another such amplitude in the diagram given in the lower panel of Fig.~\ref{fig:Diag123}. However, such contributions are at the percent level, so we do not include them in this work.} 
The diagrams in Fig.~\ref{fig:Diag123} were also found to be the dominant contributions to the nucleon's electromagnetic form factors~\cite{Cloet:2008re} and tensor charge~\cite{Xu:2015kta}, and we label them as follows: diagram 1 describes the coupling to a quark with a diquark correlation as spectator; diagram 2 describes the coupling to the quarks in a diquark correlation; and diagram 3 describes the coupling to a quark that is exchanged from one diquark correlation to another.

The PDF contributions from the diagrams in Fig.~\ref{fig:Diag123} carry different flavor weights which reflect the isospin of the quark and diquark in each diagram such that
\begin{align}
\label{eq:uv}
u_v(x) &= f^{(1)}_S(x) + \frac{1}{3}\,f^{(1)}_{A}(x) + f^{(2)}_S(x) + \frac{5}{3}\,f^{(2)}_{A}(x) \no \\
&\hs*{6mm}
 + \frac{1}{\sqrt{3}}\,f^{(2)}_{S \leftrightarrow A}(x) - \frac{4}{3}\,f^{(3)}_{A}(x) - \frac{2}{\sqrt{3}}\,f^{(3)}_{S \leftrightarrow A}(x), \allowdisplaybreaks\\[0.5em]
\label{eq:dv}
d_v(x) &= \frac{2}{3}\,f^{(1)}_{A}(x) + f^{(2)}_S(x) + \frac{1}{3}\,f^{(2)}_{A}(x) 
- \frac{1}{\sqrt{3}}\,f^{(2)}_{S \leftrightarrow A}(x) \no \\
&\hs*{13mm}
+ f^{(3)}_{S}(x) + \frac{1}{3}\,f^{(3)}_{A}(x) - \frac{1}{\sqrt{3}}\,f^{(3)}_{S \leftrightarrow A}(x),
\end{align}
where the superscript $(1,2,3)$ labels the diagrammatic mechanism, the subscripts $S$, $A$, or $S \leftrightarrow A$ label the diquark correlations involved, and $f_{S \leftrightarrow A}$ denotes the off-diagonal sum  $f_{S \rightarrow A} + f_{A \rightarrow S}$.   Note that for zero momentum transfer the operator $\sh{n}$ cannot cause a transition between a scalar and axial-vector diquark; thus, for spin-independent PDFs $f^{(2)}_{S \leftrightarrow A}(x)=0$. 

The DSE approach is formulated with Euclidean metric, analogous to lattice QCD, and therefore a direct calculation of a light-like correlation function such as $q(x)$ in Eq.~\eqref{eq:PDF_defn} is difficult, although a potential solution within the DSE approach is provided by the Nakanishi representation~\cite{Nakanishi:1969ph}. In this work we use the moment method to evaluate the diagrams in \Fig{fig:Diag123} and thereby determine the nucleon's leading-twist valence quark distribution functions. We define the moment $m$ of a PDF $q(x)$ by $\big< x^m \big>_q \equiv \int_{0}^1 dx\,  x^m \,q(x)$. In the DSE framework, any number of moments can be determined, and therefore the PDFs can be reconstructed to high precision. 

With the quark-diquark approximation made here, the explicit expression for the moments of diagram 1 is very similar to the meson case~\cite{Chang:2013pq,Chen:2016sno}, and the resulting valence quark moments for a quark of flavor $q$ are given by:
\begin{align}
\big< x^m \big>_q^{(1)} &= -\frac{1}{2\, P\cdot n} \, \sum_D\, I^{(1)}_{q,D}\, \int_k 
\lf(\frac{k \cdot n}{P \cdot n}\rg)^m \no \\ 
&\hs*{6mm}
\widebar{\Gamma}_D(p,P)\big[n \cdot \partial_k S_q(k)\big]\,\tau_D(P-k)\,\Gamma_D(p,P),
\label{eq:n_moms_diag1}
\end{align}
where $\int_k \equiv \int d^4k/(2\pi)^4$, $k$ is the momentum of the struck quark, $P$ is the nucleon momentum, $p \equiv k - P/3$ is the relative momentum between the quark and the diquark, and $S_q(k)$ is the dressed quark propagator of flavor $q$. The quantity $\Gamma_D(p,P)$ and its conjugate $\widebar{\Gamma}_D(p,P)$ are the nucleon's Faddeev amplitudes which describe the relative momentum correlation of a quark and diquark $D$, and $\tau_D$ is the effective diquark propagator. The quantities $I^{(1)}_{q,D}$ are the flavor coefficients that describe the isospin coupling weights for finding a quark of flavor $q$ in a nucleon with the configuration of diagram 1, given the isospin of the diquark configuration $D$. These coefficients can be read from Eqs.~\eqref{eq:uv} and \eqref{eq:dv}. In this work we restrict the sum over the diquark correlations $D$ to the dominant scalar ($J^P=0^+,T=0$) and axial-vector ($J^P=1^+,T=1$) diquark channels $D = S,\,A$ respectively. Finally, for the dressed vertex on the quark line we assume that the $\delta$-function in the operator can be directly applied to the dressed quark and subsequently use the Ward identity to write $S(k)\,n\cdot \mathcal{G}\,S(k) = -n \cdot \partial_k S(k)$, where $n\cdot \mathcal{G}$ is the dressed inhomogeneous vertex with driving term $i\sh{n}$. Such an approximation has proven to be an accurate representation~\cite{Nguyen:2011jy,Chang:2014lva}, as it is exact in the limit of an infrared dominant kernel~\cite{Munczek:1983dx} and preserves baryon number.

The valence quark moments for diagram 2, shown in \Fig{fig:Diag123}, are given by a loop integral over diquark momentum $K$ of an integrand having two factors: 1)  the PDF  moments for a quark within the diquark, and 2) the resulting PDF moments of the diquark as if it was an elementary constituent of the nucleon.  The first factor is 
\begin{align}
\label{eq:moms_diquarks}
\big< z^m \big>_{q/D} &= -\frac{1}{2\, K\cdot n} \,\int_k \lf(\frac{k \cdot n}{K \cdot n}\rg)^{m} \no \\
&\hs*{10mm}
\widebar{\xi}_D(\ell)\,\big[n \cdot \partial_k S_q(k)\big]\,S_{q'}(K-k)\,\xi_D(\ell),  
\end{align}
where $\ell = k - K/2$ is the relative momentum between the quarks.  We ignore a possible dependence of these moments upon  $K^2$; this is accurate because the domain of support from  the Faddeev amplitudes is heavily weighted to the deep infrared region. Therefore, $\big< z^m \big>_{q/D}$ can be factored out of the integral over $K$ and the remaining  quantity can be expressed as
\begin{align}
\label{N_moms_Diag2}
\big< y^m \big>_{D/N} &= -\frac{1}{2\, P\cdot n} \int_K \lf(\frac{K \cdot n}{P \cdot n} \rg)^{m} \no \\
&\hs*{-2mm}
\widebar{\Gamma}_D(p,P)  \, S_{q'}(P-K)\,\big[n \cdot \partial_K \tau_D(K)\big]\,\Gamma_D(p,P),
\end{align}
where the quark-diquark relative momentum is $p = 2/3\, P - K$.  Diagram 2 thus yields the final form
\begin{align}
\big< x^m \big>_q^{(2)} = \sum_D\, I^{(2)}_{q,D}\ \big< z^m \big>_{q/D} \  \big< y^m \big>_{D/N}, 
\label{Diag2_prod}
\end{align}
where the flavor factors $I^{(2)}_{q,D}$ for the diagrammatic contributions can be read from Eqs.~\eqref{eq:uv} and \eqref{eq:dv}.   For each term, this product of $m$-dependent factors is exactly what one would obtain from a convolution $q^{(2)}(x) = \int dy dz\, \delta(x - z\,y)\, q_{D}(z)\,f_{D/N}(y)$, where $q_{D}(z)$ is the quark distribution inside the diquark and $f_{D/N}(y)$ can be interpreted as the diquark light-cone distribution inside the nucleon. 

For diagram 3, shown in \Fig{fig:Diag123},  the moments are given by 
\begin{align}
\langle x^m \rangle_q^{(3)} &= -\frac{1}{2\, P\cdot n} \sum_{D',D}\, I^{(3)}_{q,D',D}\, \int_{k'} \int_k\ 
\lf(\frac{k \cdot n}{P \cdot n}\rg)^m  \no \\
&\hs*{-6mm}
\widebar{\Gamma}_{D'}(p',P) 
S_{q''}(k'')\, \tau_{D'}(P-k'')\, \xi_{D'}(\ell) \no \\
&\hs*{-6mm}
\big[n \cdot \partial_{k} S_q^T(k)\big]\,
\widebar{\xi}_D(\ell') \, S_{q'}(k')\, \tau_D(P-k')\, \Gamma_D(p,P),
\label{eq:n_moms_diag3}
\end{align}
where $k$ is the momentum of the struck quark which is exchanged between two diquark correlations, the relative momentum between the quark and diquark in the nucleon's initial and final Faddeev amplitudes are given by $p = k - P/3$ and $p' = k' - P/3$, respectively, and $\ell$, $\ell'$ are internal relative momenta of the diquark amplitudes $\xi_D$ and $\widebar{\xi}_D$.  
Momentum conservation at the vertices determines all momenta in terms of $k$ and $k'$.

\medskip
\noindent\textbf{
Faddeev Amplitudes and PDF Moment Calculations}:  For the diquark Bethe-Salpeter amplitudes $ \xi_D(\ell)$ we employ the dominant Dirac covariant in each case. Hence $\xi_S(\ell) = i \gamma_5 \, F_S(\ell^2)\, \mathcal{C} $ and $\xi_A(\ell) = \gamma_5 \gamma_\mu^T\, F_{A}(\ell^2)\, \mathcal{C}$, where the scalar functions $F_D(\ell^2)$ are given in Ref.~\cite{Cloet:2008re}, $\mathcal{C} = \gamma_2 \, \gamma_4$ is the charge conjugation operator, and superscript $T$ indicates transverse to diquark momentum $K$. The Faddeev amplitudes from Ref.~\cite{Cloet:2008re} are employed and they have the form
\begin{align}
\Gamma_D(p,P) &= G_D(p,P)\, u(P), \no \\
&= \sum_m\, i^m \, T_m(z)\, \Gamma_{D,m}(p,P) \, u(P),
\end{align}
where $p$ is the quark-diquark relative momentum, $p \cdot P = |p|\,|P|\, z$, and $T_m(z)$ are the  Chebyshev polynomials of the first kind. The Dirac matrix amplitudes have the structure
\begin{align}
\Gamma_{D,m}(p,P) = \sum_{i=1}^8 \lambda_i(p,P) \, A_{i,m}(p^2,P^2),
\end{align}
where the covariants $\lambda_i(k,P)$ are defined in Ref.~\cite{Oettel:2001kd}, with two of them applying when the diquark correlation is a scalar and six when it is in an axial-vector configuration. For example, the scalar pair are \mbox{$(\lambda_1, \lambda_2) = (1, - i \sh{p}_T) $}, while two of the axial-vector covariants are \mbox{$(\lambda_3,\lambda_4) = (P^\mu, - i \sh{p}_T \, P^\mu) $} where \mbox{$ p_T^\mu = p^\mu - P^\mu P \cdot p /P^2 $}. The conjugate Faddeev amplitude is defined by 
\begin{align}
\widebar{\Gamma}_D(p,P) &=  \bar{u}(P)\,\widebar{G}_D(p,P),
\end{align}
where \mbox{$\widebar{G}_D(p,P) = [C^{-1}\,G_D(-p,-P)\,C]^T$}. The conjugate vertex can then be expanded in a similar manner to $\Gamma_D(p,P)$ in terms of the Chebyshev polynomials of the first kind.

In Ref.~\cite{Cloet:2008re} the invariant amplitudes associated with the various quark and diquark propagators, along with the diquark Bethe-Salpeter amplitudes, were represented as phenomenological forms in terms of the entire functions $\mathcal{F}(x)=(1-{\rm e}^{-x})/x$, where $x = p^2/\omega^2$ with $p$ the momentum variable and $\omega$ the range. These parameterized forms for light quark propagators and diquark Bethe-Salpeter amplitudes are well established and well constrained by many successful studies of the associated light-meson masses and form factors~\cite{ElBennich:2010ha,Segovia:2015hra}, and they implement the dependence on the dominant mass scales of the subsystems and the leading ultraviolet power behavior.  

To consider more recent dynamical input in the present and future work, we compared these elements with direct numerical evaluations of propagators and diquark Bethe-Salpeter amplitudes from the rainbow-ladder DSE interaction of Ref.~\cite{Qin:2011dd} in the manner that has proved successful for meson properties~\cite{Chang:2013pq}. The resulting numerical elements were used here in two ways: 1) directly in computing the three PDF diagrams, and 2) as a basis for refitting the entire function forms and employing them. Our findings are that the results presented here do not change in any significant aspect. We note that Faddeev calculations~\cite{Eichmann:2016yit} of baryon masses and decays that do not employ a separable $t$-matrix approximation indicate that the separable approximation typically gives a 5\% accuracy. 

To check our calculations we employ Nakanishi-style spectral representations~\cite{Nakanishi:1969ph} of all momentum-dependent elements for the case of scalar diquarks only. This enables the use of algebraic results for Feynman integrals since all denominator factors are powers of quadratic forms multiplied by numerators that have low powers of momenta.  This approach has proved to be very efficient for simpler cases such as for meson parton distribution amplitudes~\cite{Chang:2013pq,Chang:2014lva}, parton distribution functions~\cite{Chang:2014lva,Chen:2016sno}, and form factors~\cite{Chang:2013nia,Raya:2015gva}.  For realistic nucleon calculations, the Feynman integral approach is much slower and less efficient due to the necessary numerical integration over four or five  Feynman parameter variables for every combination of the many Nakanishi fit parameters.  

In the calculation of the PDF moments, integrals of the form $f_m(P \cdot n) =  \int d^4 k\,  (k \cdot n)^m \, F(k^2, k \cdot P)$ are encountered, which naively appear divergent for large $m$. However, because of Lorentz covariance and the fact that $n^2=0$, sufficient factors of $k \cdot P$ ({\it i.e.} derivatives with respect to $k$) are contributed by $F(k^2,k \cdot P)$ to render such integrals convergent. The numerical convergence can be slow, so we adopt a method that proved effective for moments of meson distribution amplitudes~\cite{Ding:2015rkn,Li:2016dzv}: a factor $(1+ \lambda\,k^2)^{-m}$ is added to the integrand for a series of finite $\lambda$ and the results extrapolated to $\lambda \to 0$.

\begin{figure}[tbp]
\centering\includegraphics[width=\columnwidth]{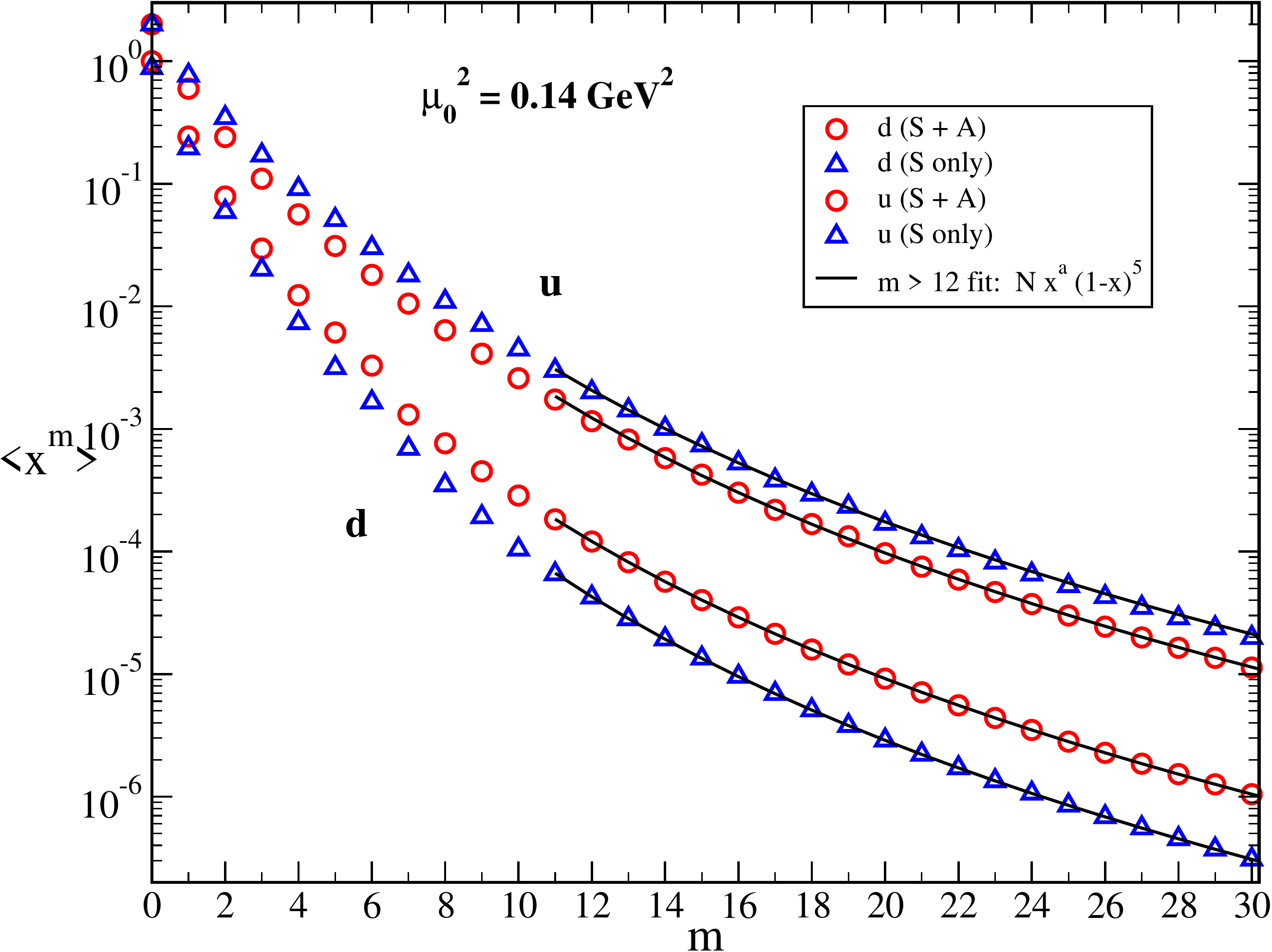}
\caption{Results for the valence $u$- and $d$-quark PDF moments $\langle x^m \rangle$ for two cases, the first where both scalar and axial-vector diquark correlations are included (open-circles), and where only scalar diquarks appear in the nucleon (open-triangles). The $m < 20$ moments are calculated by evaluating the diagrams in Fig.~\ref{fig:Diag123}, whereas the displayed moments with $m \geqslant 20$ are extrapolated by fitting the $12 \leqslant m \leqslant 19$ moments with the form $N\,x^a\,(1-x)^5$ and then extrapolating to $m \geqslant 20$. In both scenarios, $S$ only and $S+A$, baryon number is conserved.} 
\label{fig:moms}  
\end{figure}

\medskip\noindent\textbf{Results}:  Our results for the moments of the proton's leading-twist spin-independent valence quark distribution functions, $u_v(x)$ and $d_v(x)$, are illustrated in Fig.~\ref{fig:moms}. These results are obtained via numerical evaluation of the diagrams illustrated in Fig.~\ref{fig:Diag123}. Our full results, which include both scalar and axial-vector diquark correlations in the nucleon Faddeev equation, are illustrated by the open circles in Fig.~\ref{fig:moms}, whereas the open triangles correspond to a nucleon with only scalar diquark correlations. In both cases the zeroth-moments of the $u$- and $d$-quark valence PDFs are normalized so that baryon number is conserved, that is, $\big<x^0\big>_u = 2$ and $\big<x^0\big>_d = 1$.

When both scalar and axial-vector diquark correlations are present, the scalar correlation dominates by contributing 70\% of the strength.  This has important implications, {\it e.g.}, in the flavor separation of the nucleon form factors~\cite{Cates:2011pz,Cloet:2014rja} and PDFs~\cite{Cloet:2005pp}. With only scalar diquark correlations, diagram 1 dominates and contributes only to $u(x)$; diagram 2 contributes equally to $u(x)$ and $d(x)$, while diagram 3 is least important, contributing only to $d(x)$. Fig.~\ref{fig:moms} makes clear that axial-vector diquarks have a significant impact on the moments of the $d$-quark distribution function, providing significant strength at large-$x$, whereas for the $u$-quark PDF axial-vector diquarks slightly shift strength away from large-$x$. A key result seen in Fig.~\ref{fig:moms} is that the large ($ m \gtrsim 12 $) moments for both $u_v(x)$ and $d_v(x)$ -- with and without axial-vector diquarks correlations -- are very accurately described by a PDF of the form $q(x) \sim N x^a\,(1-x)^5$. This is illustrated by the solid lines which are a fit to the $12 \leqslant m \leqslant 19$ moments of the associated PDF, clearly demonstrating the preference for a $(1-x)^5$ behavior of the PDFs as $x\to 1$.

To convert our results for the PDF moments into functions of $x$ with support $0 \leqslant x \leqslant 1$ we perform a chi-squared fit of the function
\begin{align}
x\,q_v(x) = N\,x^a(1-x)^5\lf(1 + b\,\sqrt{x} + c\,x\rg)
\label{eq:pdfs}
\end{align}
to the moments given in Fig.~\ref{fig:moms}, and since the large-$x$ behavior clearly favors $(1-x)^5$ we fix this exponent beforehand. A fit to the moments of $u_v(x)$ gives $N=30.59,\,a=2.510,\,b=15.52,\,c=-11.26$ and for $d_v(x)$ we obtain $N=7.071,\,a=1.7887,\,b=14.33,\,c=-14.42$. Because the higher PDF moments are much smaller than the leading moments this parametrization is accurate on the domain $0.2 \lesssim x \lesssim 0.7$, and as we will see, an accurate determination of  the large-$x$ behavior requires a more careful analysis of the higher moments. 

\begin{figure}[tbp]
\centering\includegraphics[width=\columnwidth]{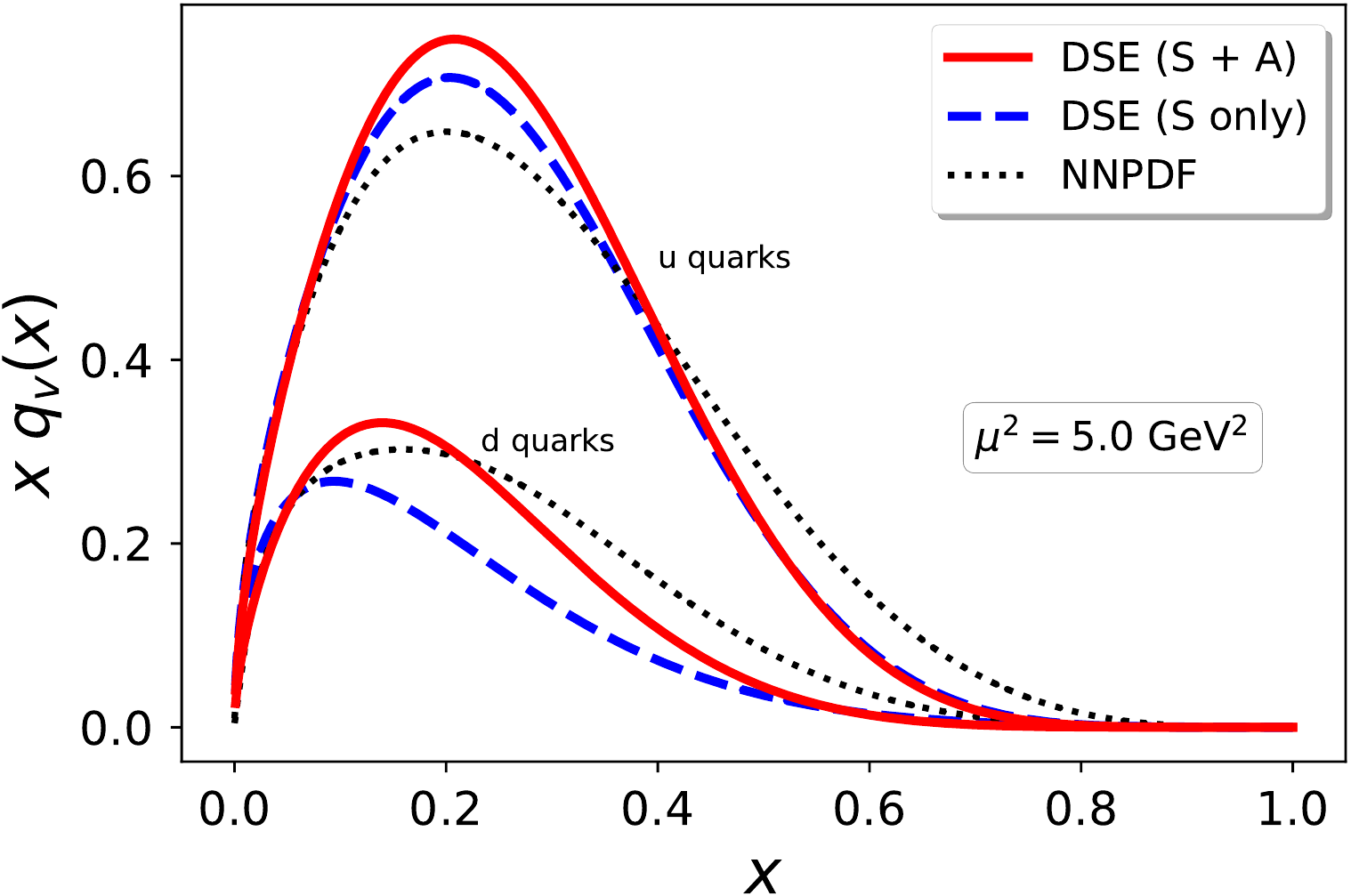}
\caption{Results for the $u$ and $d$ valence quark distributions in the proton, evolved from the model scale of $\mu^2_0 = 0.14\,$GeV$^2$, to $\mu^2 = 5.0\,$GeV$^2$ using NLO DGLAP evolution~\protect\cite{Botje:2010ay,Botje:2016wbq}. The solid curve is our full result which includes both scalar and axial-vector diquarks, and the dashed curve is our result with only scalar diquarks. The dotted curve is the empirical NLO result from NNPDF 3.0~\protect\cite{Ball:2012cx}.}
\label{fig:PDFs_Nov1_5GeV2}  
\end{figure}

Most models of hadron PDFs do not incorporate a specific dependence upon resolving scale; hence the inherent model scale  must be deduced outside of the model. Here the model scale $\mu_0^2$ is determined so as to produce the best fit of the valence $u$-quark distribution, after DGLAP evolution, to the empirical NNPDF 3.0 NLO results~\cite{Ball:2014uwa} at a scale of $\mu^2 = 5\,$GeV$^2$. This gives  $\mu_{0}^2 = 0.14$~GeV$^2$.  Our results for $u_v(x)$ and $d_v(x)$ at a scale of $\mu^2 = 5\,$GeV$^2$ are shown in \Fig{fig:PDFs_Nov1_5GeV2} and compared to the NNPDF 3.0 NLO empirical parametrization~\cite{Ball:2014uwa}. We find quite reasonable agreement.  A clear outcome is that the $d$ quark is softer than the $u$ quark due to the 70\% dominance of scalar diquark correlations which isolate the $d$ quark inside a light scalar diquark. At large $x$ our results lack strength when compared to the empirical parametrizations.   This is a consequence of the $(1-x)^5$ behavior of our results as $x \to 1$, whereas the empirical results are closer to the QCD counting rule expectation of $q(x) \stackrel{x\to 1}{\simeq} (1-x)^3$~\cite{Drell:1969km,West:1970av,Brodsky:1974vy}.

Our large-$x$ behavior can be understood by first considering the relation $\mathcal{F}_2(x) \sim Q^2 |F_1(Q^2)|^2$, derived in Ref.~\cite{Farrar:1975yb} and valid for very large $Q^2$ with fixed $W^2 \sim Q^2\, (1-x)$. Here, $\mathcal{F}_2(x)$ is the nucleon's inclusive scattering structure function and $F_1(Q^2)$ is its elastic electromagnetic form factor. In QCD one expects the behavior $F_1(Q^2) \sim (Q^2)^{-n_s}$, where $n_s$ is the minimum number of spectator partons, because the large $Q$ initially transferred to one quark from the incident photon can only connect a three-quark state having soft internal relative momenta to another such state through successive gluon exchanges between a minimum of $n_s$ quark pairs. The linking quark propagators do not scale with asymptotic $Q$ due to the mass-shell constraint on the final nucleon state. Therefore, the earlier relation leads to the  Drell-Yan--West relation~\cite{Drell:1969km,West:1970av} $\mathcal{F}_2(x) \sim (1-x)^{2 n_s -1}$ which implies $q(x) \stackrel{x\to 1}{\simeq} (1-x)^3$. 

Key to obtaining this result is that the $qq$ correlation $t$-matrix is given by one-gluon exchange in the ultraviolet limit of QCD.  However, in our calculation we use a separable approximation for the $qq$ $t$-matrix, which always introduces a product of two Bethe-Salpeter amplitudes, and therefore the separable $t$-matrix scales as $1/(q_f^2 \, q_i^2)$ rather than as a single gluon propagator $1/(q_f - q_i)^2$ where $q_f, q_i$ are the final and initial relative momenta of the $qq$ pair. Since the final correlated pair involved in mitigating the high $Q$ can have a soft final state relative momentum, the net result in the separable case is that $\mathcal{F}_2(x) \sim (1-x)^{4 n_s -3}$. Therefore, the separable approximation $qq$ $t$-matrix necessarily leads to a large-$x$ behavior of $q(x) \stackrel{x\to 1}{\simeq} (1-x)^5$. We therefore find the result that, while being reliable for numerous observables~\cite{Cloet:2008re}, the quark-diquark approximation for the nucleon necessarily breaks down for PDFs in the extreme ultraviolet limit associated with $x \to 1$. Nevertheless, results for PDF ratios can be considered reliable in this limit because the leading power behavior cancels in the ratio.

\medskip
\noindent\textbf{$\vect{d/u}$ Ratio at Large-$\vect{x}$:} Extracting the $x\to 1$ behavior of the PDFs requires a careful analysis of the high moments.   In particular we find that at least five $\big<x^m\big>_q$ moments with $m \geqslant 12$ are required to accurately extract the exponent and normalization of the PDFs as $x \to 1$. To determine the $d(x)/u(x)$ ratio on the domain $0.9 \lesssim x \leqslant 1$ we accurately fit the $\langle x^m \rangle_q$ moments for $m = 12,\ldots,19$ to the function $q(x) \simeq N\, x^a\,(1-x)^b$, whose moments are proportional to the Euler beta function $B(1+a+m,1+b)$.  This fit indicates a clear preference for $b=5$. This is true not only for the $u$ and $d$ PDFs overall but also for each component of the diagrams that appear in Eqs.~\eqref{eq:uv} and \eqref{eq:dv}, except for  $f^{(3)}_{S \leftrightarrow A}(x)$. Therefore, contrary to naive scalar diquark dominance models~\cite{Close:1973xw}, 
we do not find a power suppression of the $d$-quark distribution relative to the $u$-quark distribution  as $x\to 1$; for the $d/u$ ratio with only scalar diquarks, we find the small but finite result $d(x)/u(x) \stackrel{x\to 1}{=} 0.011 \pm 0.003$, where the error reflects numerical uncertainties in the moments of the PDFs and the fit procedure.

As  Fig.~\ref{fig:moms} demonstrates, axial-vector correlations have a dramatic impact on the high moments of the $d$-quark PDF.    The aforementioned fit to the moments for the complete scalar and axial-vector calculation gives our full result for the $d/u$ ratio:
\begin{align}
d(x)/u(x) \stackrel{x\to 1}{=} 0.087 \pm 0.010,
\end{align}
which is about eight times larger than the pure scalar diquark result. Again, the error reflects numerical uncertainties in the moments of the PDFs and the fit procedure. The value of the $d/u$ ratio in the limit $x\to 1$ is scale invariant, and our DSE result compares well with the most recent analysis from the CTEQ--Jefferson Lab collaboration (CJ15)~\cite{Accardi:2016qay}, which find $d/u=0.09\pm 0.03$. We note that this present DSE result for the direct calculation of the $d(x)/u(x)$ ratio as $x\to 1$ takes a value roughly one-third of that of an earlier DSE estimate based on a flavor breakdown of the proton's elastic Dirac form factor obtained at the infrared dominant point $Q^2=0$~\cite{Roberts:2013mja,Xu:2018eii}.

\begin{figure}[tbp]
\centering\includegraphics[width=\columnwidth]{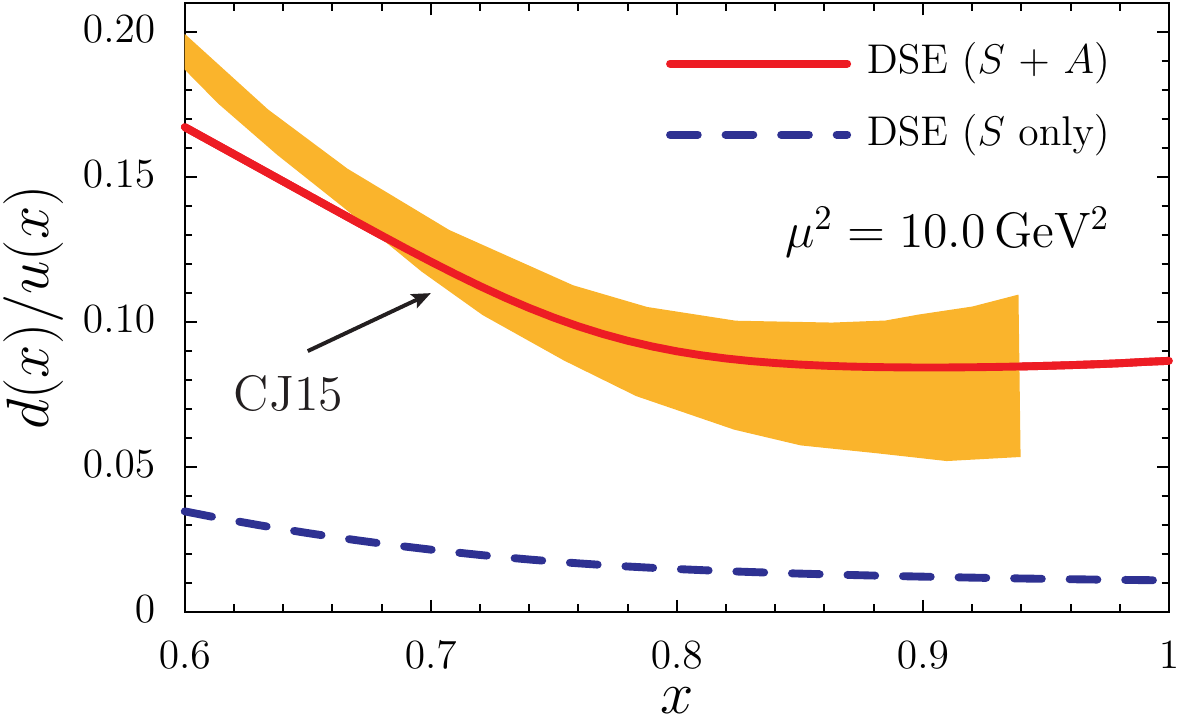}
\caption{DSE result for $d(x)/u(x)$ at a scale of $\mu^2 = 10\,$GeV$^2$. The dashed curve is the scalar diquark only result and our full result including axial-vector diquark correlations is given by the solid curve. The shaded area is the CJ15 empirical result~\cite{Accardi:2016qay}. Note the DSE scalar diquark only result does not vanish as $x \to 1$, contrary to naive model 
expectations~\cite{Close:1973xw}.   }
\label{fig:doveru}  
\end{figure}

\looseness=-1
In Fig.~\ref{fig:doveru} we present our results for $d(x)/u(x)$ over the domain $0.6 \leqslant x \leqslant 1$ at a scale of $Q^2 = 10\,$GeV$^2$ and compare them with the CJ15 result~\cite{Accardi:2016qay}.  To construct the ratio in this domain, we have taken the result of the fit of Eq.~\eqref{eq:pdfs} to  all moments, and used the produced ratio at $x \lesssim 0.7$ with smooth  interpolation to the  $x \gtrsim 0.9$ result produced by the specialized fit to  the $m \gtrsim 12$ moments.   Again, for our full $S+A$ result, we find excellent agreement with CJ15 for $x \gtrsim 0.7$.   The flattening of the $d/u$ ratio above $x \sim 0.8$ reflects the increasing dominance of the $(1-x)^5$ component, and this factor cancels increasingly in the ratio as $x\to 1$.

\medskip
\noindent\textbf{Anti-quark and Gluon Momentum Fractions:}  To assess how these valence quark distribution function results might fit into a broader view of the nucleon's PDFs, we estimate the light-cone momentum fraction carried by sea-quarks at the model scale through a simple pion cloud model. A key feature of the DSE approach is that it contains explicit gluon degrees of freedom, which give rise to {\it inter alia} running quark masses and (relative) momentum dependent Bethe-Salpeter and Faddeev amplitudes. For PDFs at the model scale this implies that the dressed quarks do not carry 100\% of the light-cone momentum of the nucleon;  the remainder is carried by gluons. The DSE approach is also Poincar\'e-covariant and respects momentum conservation; therefore we use the momentum sum rule to estimate the momentum fraction carried by  gluons. For a nucleon composed of only valence quarks and with only scalar quark-quark correlations, we find $\big<x\big>_g = 0.015$; thus gluons carry only 1.5\% of the nucleon's light-cone momentum. However, for our full result which also includes axial-vector quark-quark correlations, we obtain $\big<x\big>_g = 0.16$, and hence gluons carry 16\% of the light-cone momentum at the model scale  $\mu_0^2 = 0.14\,$GeV$^2$. These results are summarized in the first two rows of Tab.~\ref{tab:moms}, together with the light-cone momentum fractions carried by the quarks.

\looseness=-1
It has long been known that the observed flavor asymmetry of the sea-quarks -- at least for low moments -- is qualitatively explained by the pion cloud mechanism~\cite{Thomas:1983fh,Steffens:1996bc}. More generally, within a meson-baryon virtual fluctuation model that is phenomenologically successful~\cite{Alwall:2005xd}, the pion cloud mechanism introduces Fock space components $\pi^\alpha + N^\alpha$ to the proton with probability $Z$, and leads to quark distribution moments given by
\begin{align}
\big< x^m \big>_q &= (1-Z)\,\big< x^m \big>_{q/p} \no \\
&\hs*{9mm}
+ Z \sum_\alpha I_\alpha \left[ \langle x^m \rangle_{q/\pi^{\alpha}/p} + \langle x^m \rangle_{q/N^\alpha/p} \right],
\label{eq:q_pi_cloud}
\end{align}
where $\langle x^m \rangle_{q/V/p}$ is the quark PDF moment within the proton's virtual component $V = \pi^\alpha, N^\alpha$, which in line with the convolution picture we approximate by $\langle x^m \rangle_{q/V/P} = \langle x^m \rangle_{q/V} \, \langle x^m \rangle_{V/P}$. All pion and nucleon quantities that enter the right-hand side of Eq.~\eqref{eq:q_pi_cloud} are ``bare'' and can be interpreted as quark-core contributions to the pion and nucleon states, and here we include the $\pi^0$--$\,p$ and $\pi^+$--$\,n$ components of the proton which enter with isospin probabilities of $I_0 = \frac{1}{3}$ and $I_1 = \frac{2}{3}$, respectively. 

\begin{table*}[tbp]
\addtolength{\tabcolsep}{7.5pt}
\addtolength{\extrarowheight}{2.2pt}
\begin{tabular}{l|ccccccccccccccccccc}
\hline\hline
model      & $\mu^2$ & $\big<x^0\big>_u$ & $\big<x^0\big>_{\bar{u}}$ & $\big<x^0\big>_d$ & $\big<x^0\big>_{\bar{d}}$ & $\big<x\big>_u$ & $\big<x\big>_{\bar{u}}$ & $\big<x\big>_d$ & $\big<x\big>_{\bar{d}}$ & $\big<x\big>_g$     \\[0.2em]
\hline
$S$        & $\mu_0^2$      &  2      &  0       &  1      & 0       & 0.764  &  0        &  0.221  & 0      &  0.015   \\
$S+A$      & $\mu_0^2$      &  2      &  0       &  1      & 0       & 0.596  &  0        &  0.243  & 0      &  0.161   \\
$S+A+\pi$  & $\mu_0^2$      &  2.034  &  0.034   &  1.168  & 0.168   & 0.550  &  0.0025   &  0.278  & 0.0124 &  0.156   \\
$S+A+\pi$  & $5.0\,$GeV$^2$ &  $-$    &  $-$     &   $-$   &  $-$    & 0.265  &  0.0244   &  0.145  & 0.0289 &  0.478   \\
NNPDF 3.0  & $5.0\,$GeV$^2$ &  $-$    &  $-$     &  $-$    &  $-$    & 0.307  &  0.0302   &  0.148  & 0.0362 &  0.444    \\
\hline\hline
\end{tabular}
\caption{Results for the $m=0,\,1$ parton moments  from the valence only calculation (first two rows) and from the extension via the pion cloud model (third row) at the model scale $\mu_0^2 = 0.14\,$GeV$^2$.   The gluon light-cone moment fractions are inferred from the  momentum sum rule.
The fourth row presents our full results evolved to $\mu^2 = 5\,$GeV$^2$ to be compared to the NNPDF 3.0~\cite{Ball:2014uwa} empirical results in the fifth row. We do not give $m=0$ results for the evolved PDFs because the small-$x$ behavior of these distributions under DGLAP evolution make this difficult to determine. For PDFs at $\mu^2 = 5\,$GeV$^2$ a small amount of light-cone momentum is also carried by $s +\bar{s} + c + \bar{c}$, which is generated by the DGLAP evolution equations.}
\label{tab:moms}
\end{table*}

In this simple pion-cloud model the sea-quark asymmetry is given by $\big< x^0 \big>_{\bar d} - \big< x^0 \big>_{\bar u} = \frac{2}{3}\,Z$, and therefore any finite $Z$ produces an asymmetry. To constrain $Z$ we use the Gottfried sum rule~\cite{Gottfried:1967kk} result from Ref.~\cite{Abbate:2005ct}:
\begin{align}
S_G &= 0.244 \pm 0.045
\simeq \frac{1}{3} - \frac{2}{3}\lf[\big< x^0 \big>_{\bar d} - \big< x^0 \big>_{\bar u}\rg],
\end{align}
which by taking the central value implies $Z = 0.20$.\footnote{The empirical Gottfried sum rule result used here is valid for $1.5 \leqslant Q^2 \leqslant 4.5\,$GeV$^2$, and this scale evolution should be taken into account when determining $Z$. However, given the qualitative nature of the pion cloud model we ignore this evolution, however in practice it would slightly increase our value for $Z$.} 
Therefore the probability of finding a bare proton, without a pion cloud, is 80\%. Results for the zeroth moments of the PDFs then take the simple form: $\big< x^0 \big>_u = 2 + \frac{1}{6}\,Z$, $\big< x^0 \big>_d = 1 + \frac{5}{6}\,Z$, $\big< x^0 \big>_{\bar{u}} = \frac{1}{6}\,Z$, and $\big< x^0 \big>_{\bar{d}} = \frac{5}{6}\,Z$. These expressions clearly satisfy baryon number conservation, and our numerical results are given in the third row of Tab.~\ref{tab:moms}. For the light-cone momentum fractions this simple pion cloud model gives
\begin{align}
\big< x \big>_u &= (1-Z)\,\big< x \big>_{u/p} \no \\
&\hs*{0mm}
+ \tfrac{1}{3}\,Z\lf[\langle x \rangle_{u/p} + 2\,\langle x \rangle_{d/p}\rg]\lf(1 - \tfrac{m_\pi}{m_N}\rg)
+ \tfrac{5}{12}\,Z\,\tfrac{m_\pi}{m_N},  \\[0.3em] 
\big< x \big>_d &= (1-Z)\,\big< x \big>_{d/p} \no \\
&\hs*{0mm}
+ \tfrac{1}{3}\,Z\lf[\langle x \rangle_{d/p} + 2\,\langle x \rangle_{u/p}\rg]\lf(1 - \tfrac{m_\pi}{m_N}\rg)
+ \tfrac{1}{12}\,Z\,\tfrac{m_\pi}{m_N}, \allowdisplaybreaks \\[0.3em]
\big< x \big>_{\bar{u}} &= \tfrac{1}{12}\,Z\,\tfrac{m_\pi}{m_N}, \hs*{15mm}
\big< x \big>_{\bar{d}} = \tfrac{5}{12}\,Z\,\tfrac{m_\pi}{m_N},
\end{align}
where, as in Ref.~\cite{Alwall:2005xd}, we take $\langle x \rangle_{\pi^0/p} \simeq \langle x \rangle_{\pi^+/p} = \frac{m_\pi}{m_N}$ and $\langle x \rangle_{p/p} = \langle x \rangle_{n/p} = 1 - \frac{m_\pi}{m_N}$, and assume that the dressed quarks carry all the momentum of the pion. Our numerical results for the light-cone momentum fractions at the model scale of $\mu_0^2 = 0.14\,$GeV$^2$ are given in the third row of Tab.~\ref{tab:moms}. The fourth row of Tab.~\ref{tab:moms} presents the model scale results from row four, evolved using NLO DGLAP evolution to the scale $\mu^2 = 5.0\,$GeV$^2$, and these are to be compared to the empirical fit to experiment in the fifth row~\cite{Ball:2014uwa}.   The simple pion cloud model, coupled with our DSE results, gives a qualitatively good account of the light-cone momentum fractions carried by the quarks and gluons in the nucleon.

\medskip
\noindent\textbf{Summary}: The core features of the leading-twist spin-independent valence quark distribution functions of the proton are -- for the first time -- determined within a realistic DSE framework. Importantly the nucleon Faddeev amplitudes contain contributions from the dominant scalar and axial-vector diquark correlations with momentum-dependent Bethe-Salpeter structure. The $x$-dependence of the PDFs is extracted from twenty moments calculated within a rainbow-ladder truncation of the DSEs of QCD. The DSE framework used here is the same as that employed in Ref.~\cite{Cloet:2008re} to calculate and predict the nucleon's electromagnetic form factors. We find that scalar diquark correlations are dominant but that axial-vector quark-quark correlations still have a dramatic impact on the behavior of the PDFs, particularly for the $d$-quark distribution from moderate-to-large values of $x$. We compare our DSE results with the empirical NNPDF parameterizations at a scale of $\mu^2 = 5\,$GeV$^2$, finding good agreement.  We conclude that the present DSE framework has captured the various mass-scales that dominate the underlying QCD dynamics for leading-twist spin-independent PDFs.

The large-$x$ behavior of the PDFs is analyzed in detail. We find that the PDFs on the domain $0.9 \lesssim x \leqslant 1$ can be well constrained if the $m=12,\ldots,19$ moments are accurately known, and in the  $x\to 1$ limit we make the prediction that the $d/u$ ratio takes the value $d(x)/u(x) \stackrel{x\to 1}{=} 0.087 \pm 0.010$. This result includes contributions from both scalar and axial-vector diquark correlations in the nucleon Faddeev amplitude. If only scalar diquark correlations are included, we still find that each of the three dominant diagrammatic mechanisms yield the same leading power-law behavior as $x \to 1$, and therefore with only scalar diquarks we find a small, finite result for the end-point $d/u$ ratio, namely, $d(x)/u(x) \stackrel{x\to 1}{=} 0.011 \pm 0.003$. This result is in contrast to the standard scalar diquark dominance argument which suggests zero for this ratio at $x\to1$~\cite{Close:1973xw}. These DSE results now provide a foundation from which to pursue investigations of higher-twist and spin-dependent PDFs, along with transverse momentum dependent parton distributions.

\smallskip
\noindent\textit{Acknowledgments:}
We are grateful to Gerald A. Miller and Anthony W. Thomas for helpful comments, and KB acknowledges support from the graduate student visitor program of Argonne National Laboratory which enabled several beneficial visits. This work was supported by the U.S. Department of Energy, Office of Science, Office of Nuclear Physics, contract no. DE-AC02-06CH11357; by the National Science Foundation, grant no.\ NSF-PHY1516138; and the Laboratory Directed Research and Development (LDRD) funding from Argonne National Laboratory, project no. 2016-098-N0 and project no. 2017-058-N0.


%

\end{document}